\documentclass[3p,times,twocolumn]{elsarticle}

\usepackage{ecrc}

\volume{00}

\firstpage{1}

\journalname{Nuclear Physics B Proceedings Supplement}

\runauth{}

\jid{nuphbp}

\jnltitlelogo{Nuclear Physics B Proceedings Supplement}

\usepackage{amssymb}

\biboptions{compress}

\usepackage[figuresright]{rotating}

\begin{document}

\begin{frontmatter}



\dochead{}

\title{$\alpha_{s}$ from the (revised) ALEPH data for $\tau$ decay}


\author{Diogo Boito$^a$, Maarten Golterman$^b$, Kim Maltman$^c$, James Osborne$^d$ and Santiago Peris$^{e,1}$}

\address{$^a$Physik Dept.  T31,
Technische Univ. M\"{u}nchen, D-85748 Garching, Germany\\
$^b$Dept. of Physics and Astronomy,
San Francisco State University, San Francisco, CA 94132, USA\\
$^c$Dept.  of Mathematics and Statistics, York Univ., Toronto, ON CANADA M3J 1P3 and\\
CSSM, Univ. of Adelaide, Adelaide, SA 5005 AUSTRALIA\\
$^d$Physics Dept., Univ. of Wisconsin-Madison, Madison, WI 53706, USA\\
$^e$Dept. of Physics, Universitat Aut\`onoma de Barcelona,  E-08193 Bellaterra, Barcelona, Spain}

\fntext[label1]{Speaker}

\vspace{-1cm}
\begin{abstract}
We present a new analysis of $\alpha_s$ from hadronic $\tau$ decays based on the recently revised ALEPH data. The analysis is based on a strategy which we previously applied to the OPAL data. We critically compare our strategy to the one  traditionally used and comment on the main differences. Our analysis yields the values $\alpha_s(m_\tau^2)=0.296\pm 0.010$
using fixed-order perturbation theory, and $\alpha_s(m_\tau^2)=0.310\pm 0.014$
using contour-improved perturbation theory.   Averaging these values with
our previously obtained values from the OPAL data, we find
$\alpha_s(m_\tau^2)=0.303\pm 0.009$, respectively, $\alpha_s(m_\tau^2)=0.319\pm 0.012$, as the most reliable results for $\alpha_s$ from $\tau$ decays currently available.
\end{abstract}


\end{frontmatter}



\begin{figure}[h]
\includegraphics*[width=3.8cm]{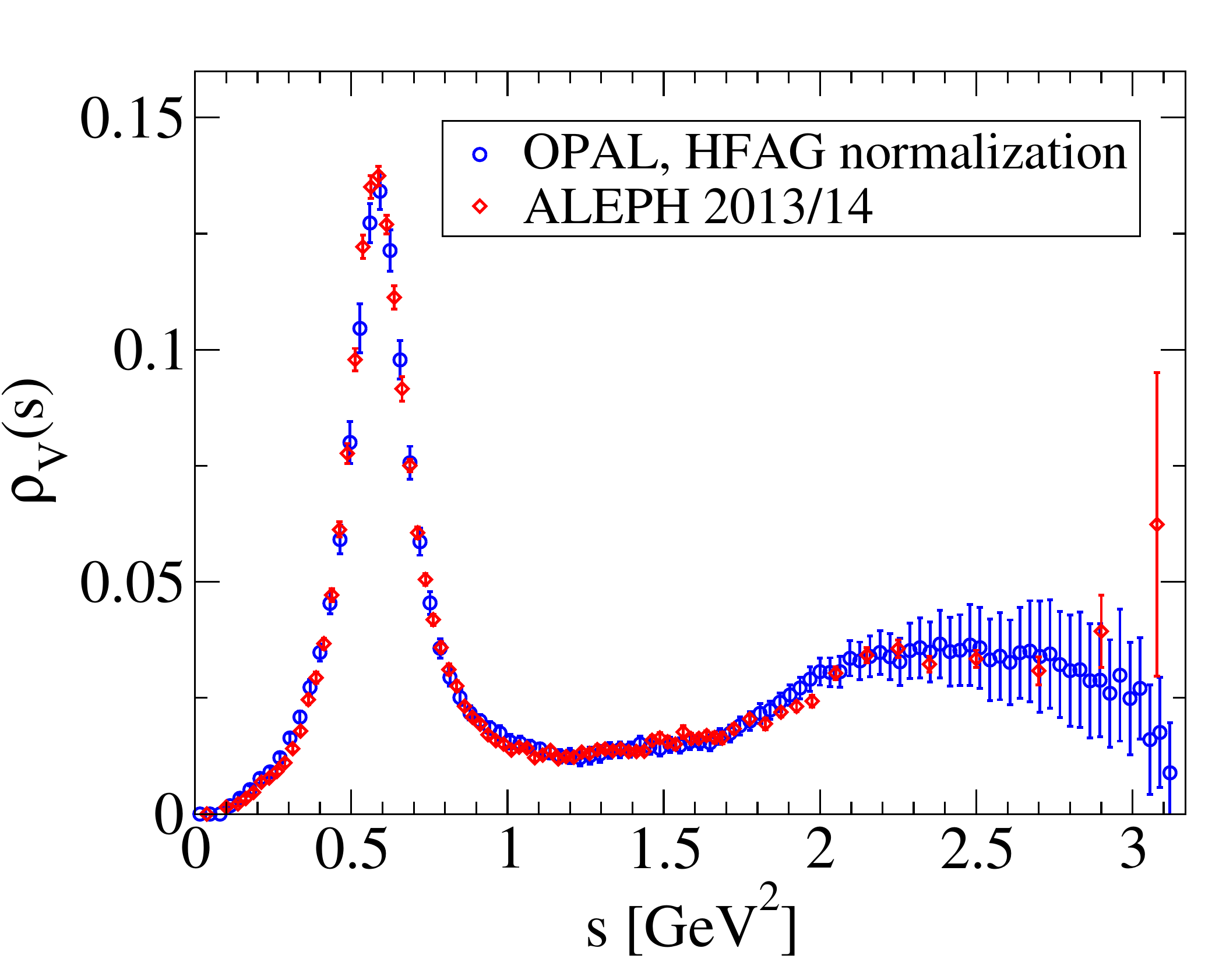}
\includegraphics*[width=3.8cm]{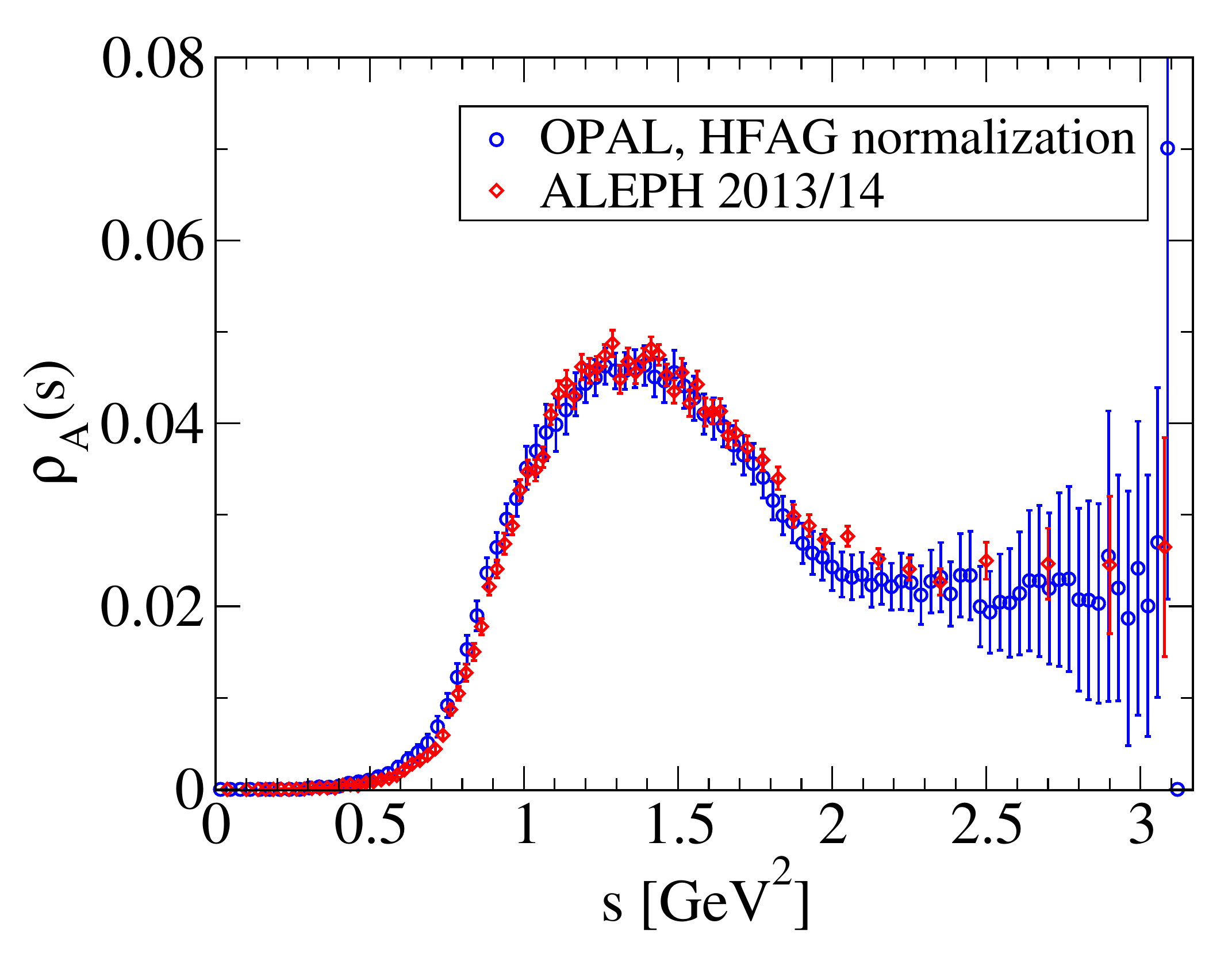}
\vspace{-.2cm}
\caption{\label{fig.1} ALEPH (red) and OPAL (blue) spectral functions. Left panel:
$I=1$ vector channel; right panel: $I=1$ continuum (pion-pole subtracted)
axial channel.}
\end{figure}

Thanks to the work by ALEPH \cite{ALEPH13} and OPAL \cite{OPAL}, we have at our disposal the vector (V) and axial-vector (A) hadronic spectral functions, $\rho_{V/A}(s)$, for $s \leq m^2_{\tau}$. They are shown in Fig. \ref{fig.1}.  The new ALEPH data in \cite{ALEPH13} supersede the previous ALEPH data, correcting a problem with the data correlation matrices discovered in Ref.~\cite{Tau10}. The revised data pass the test used in Ref.~\cite{Tau10} to uncover the original problem. We have now reanalyzed the revised data using the method previously applied to the OPAL data \cite{alphas1,alphas2}.

Although there is, in general, good agreement between the ALEPH and OPAL data, there are some points of difference, as can be seen in Fig.~\ref{fig.1}. The shoulder seen in the V channel by the OPAL data for $s\simeq 3\ \mathrm{GeV}^2$  is not seen by ALEPH, whose data stay higher than OPAL's, even though the errors are so large that is difficult to make any definitive conclusions. Additional differences exist in the regions  $\lesssim 0.5$~GeV$^2$ and around $2$~GeV$^2$ in the vector channel, with possibly anti-correlated tensions in the same regions in the axial channel. Since these data are being used for the extraction of a fundamental parameter in QCD, i.e., $\alpha_s$, it is important that they are as precise and reliable as possible.  New data collected by Belle and Babar could be extremely useful in this regard, and we urge these collaborations to produce their own inclusive V and A spectra.

Analyticity, together with Cauchy's theorem, allows us to relate the spectral function to the vacuum polarization as
\begin{eqnarray}
\label{cauchy}
&&\hspace{-1.5cm}I^{(w)}_{V/A}(s_0)\equiv\int_0^{s_0}\frac{ds}{s_0}\,w(s)\,\rho^{(1+0)}_{V/A}(s)\nonumber \\
&&\vspace{-.5cm}=-\frac{1}{2\pi i}\oint_{|s|=s_0}
\frac{ds}{s_0}\,w(s)\,\Pi^{(1+0)}_{V/A}(s)\ .
\end{eqnarray}
This equation is valid for any $s_0>0$ and any weight $w(s)$ analytic inside and on the contour \cite{shankar,BNP}. In Eq. (\ref{cauchy}) we have restricted our attention to the kinematic-singularity-free $1+0$ spin combination. We emphasize that Eq.~(\ref{cauchy}) is exact  when  the exact $\Pi(s)$  is used. However, this is not what is being done in the standard analysis \cite{ALEPH13,OPAL}. Instead, one assumes that the replacement  $ \Pi(s) \rightarrow \Pi_{\rm OPE}(s) $, where $\Pi_{\rm OPE}$ is the Operator Product Expansion of $\Pi(s)$,  is a sufficiently accurate approximation. Since $\Pi_{\rm OPE}(s) $ contains, besides the usual condensates, also the perturbative series,  this replacement in Eq.~(\ref{cauchy}) allows the extraction of $\alpha_s$ if sufficiently accurate data are available for $\rho(s)$. The difference $\Pi(s) - \Pi_{\rm OPE}(s) $ represents, by definition, the duality violating (DV) part, $\Pi_{DV}(s)$, of $\Pi(s)$. In the standard analysis, $\Pi_{DV}(s)$ is assumed to be negligible.

However, how do we know that  $\Pi_{DV}(s) $ is really small enough? Even if we accept that $s\sim m_\tau^2$ is large
enough that higher dimension OPE condensate contributions to $\Pi (s)$ are relatively small, the contour in Eq. (\ref{cauchy}) includes points which lie arbitrarily close to the Minkowski axis, where we know that the OPE must fail badly since its imaginary part does not resemble anything like the spectrum.\footnote{This behavior of the OPE, where the convergence in the complex plane depends on the angle, is typical of an asymptotic expansion. } One may try to minimize the contribution from this dangerous  region around the Minkowski axis by choosing a polynomial $w(s)$ in Eq.~(\ref{cauchy}) with a high-order zero at $s=s_0$, i.e., with a high-order ``pinching" in the weight $w(s)$ \cite{DP1992}. However, such a high-order zero necessarily requires a high-order polynomial and, through Eq. (\ref{cauchy}), this brings in the unwelcome contribution of unknown high-dimension condensates.  Furthermore, since the OPE is expected to be an asymptotic expansion, there is no reason for the contribution from these high-dimension condensates to be small and, in any case, any assumption on their size should be tested.  Pinching can only reduce the contribution from DVs by increasing the dimension of the condensates contributing to Eq.~(\ref{cauchy}). The bottom line is that  it is not possible to \emph{simultaneously} reduce the effect of duality violations and the contribution from high-dimension condensates through pinching. However,  in the standard analysis, pioneered in Refs.~\cite{BNP,DP1992} and leading to the results of Refs.~\cite{ALEPH13} and \cite{OPAL},  this is precisely what is done: The contribution from DVs are ignored altogether while, at the same time, condensates contributing to Eq.~(\ref{cauchy}) of dimension 10 and higher are also set to zero.  In Ref.~\cite{Boito:2014sta}, we have performed a number of tests of the self-consistency of these  assumptions and found they are not supported by the data (see also \cite{MY08,Boito:2012mr}).

In order to be able to quantify the contribution from DVs we have, in the absence of a theory of duality violations, employed the following physically-motivated parametrization \cite{CGP05,CGPmodel,CGP,russians}
\begin{equation}
\label{ansatz}
\rho_{V/A}^{\rm DV}(s)=
e^{-\delta_{V/A}-\gamma_{V/A}s}\sin{(\alpha_{V/A}+\beta_{V/A}s)}\ ,
\end{equation}
which we expect to be valid for large enough $s$. We refer to Ref.~\cite{Boito:2014sta} and references therein for details. Suffice it to say here that this functional form is based on large-$N_c$ and Regge theory phenomenological considerations, as well as general properties of asymptotic expansions.

The form in Eq.~(\ref{ansatz}) introduces 4 new parameters per channel to be determined by the data. In this language, the standard analysis  corresponds to the \emph{choice} $\delta_{V,A} \rightarrow \infty$. In our analysis, on the contrary, we may now take DVs explicitly into account, and modify the second line of Eq. (\ref{cauchy}) to read \cite{CGP}
\begin{equation}
\label{sumrule}
 \hspace{-1cm}-\frac{1}{2\pi i }\oint_{|s|=s_0}
\hspace{-.15cm}\frac{ds}{s_0}\,w(s)\,\Pi^{(1+0)}_{V/A,{\rm OPE}}(s)-
\hspace{-.1cm}\int_{s_0}^\infty \hspace{-.15cm}\frac{ds}{s_0}\,w(s)\, \rho_{V/A}^{\rm DV}(s).
\end{equation}
Using experimental data for $\rho^{(1+0)}_{V/A}$ in the first line of Eq. (\ref{cauchy}), whose RHS we will refer to as $I^{(w)}_{\mathrm{exp}}(s_0)$, and equating it to the expression in Eq. (\ref{sumrule}), which we will refer to as   $I^{(w)}_{\mathrm{th}}(s_0)$, one obtains the master equation to be used in determining all the DV parameters,  $\alpha_s$ and the condensates,  for $s_0$ in a window $s_{\mathrm{min}} \leq s_0 \leq m^2_{\tau}$ and a judiciously chosen set of weights, $w(s)$.

 Let us now compare the main differences between the standard analysis and our new strategy \cite{Boito:2014sta}.

In the standard analysis \cite{DP1992,ALEPH13,OPAL} one
\begin{itemize}
  \item uses 5 pinched weights in the sum rule of Eqs. (\ref{cauchy}-\ref{sumrule}): $w_{k \ell}(y)=(1-y)^{2+k} (1+2y) y^\ell$, with  $y=s/s_0$ and $(k,\ell)=(0,0),(1,0),(1,1),(1,2),(1,3)$,
   \item neglects DVs, i.e., one sets $\delta_{V,A}\rightarrow \infty$ in Eq. (\ref{ansatz}),
  \item performs a fit for the 4 parameters $\alpha_s(m^2_{\tau})$ and the dimension$-D$ condensates, $C_D$, with $D={4,6,8}$ using the $s_0=m_{\tau}^2$ values of the 5 weighted spectral integrals only,
       \item sets the other OPE condensates, $C_{10,12,14,16}$ to zero by fiat, even though they contribute to Eq.~(\ref{cauchy}) for the above weights at leading order in $\alpha_s$,
  \item may use the V and A channels, but assumes V+A to be more reliable. This is done even though the V channel leads to a fit  with a  much better $\chi^2$ than the one for the V+A channel \cite{ALEPH13}. The V, A and V+A channel fits are also found to yield inconsistent values of the $D=4$ gluon condensate,
  \item does several  checks on the analysis such as, e.g.,  the verification of the Weinberg sum rules (WSRs)~\cite{SW}. The WSRs are not satisfied if $\rho_{DV}(s)$ is neglected above $s=m_\tau^2$.

\end{itemize}
Note that DVs are, as is well known, clearly present in $\rho (s)$ in the region of $s$ accessible in hadronic $\tau$ decays.

This is to be compared to the new strategy proposed in Refs.~\cite{alphas1,alphas2,Boito:2014sta}. There one
\begin{itemize}
  \item avoids the use of weights $w(y)$ with a linear term in $y$. Ref.~\cite{BBJ12} found perturbation theory for these weights to be unreliable,
  \item does not assume any condensate contributions to vanish but, instead,  lets the data determine them,
  \item does not assume DVs to vanish but, again,  lets the data determine the parameters in Eq. (\ref{ansatz}),
  \item makes fits to the data with only the 3 weights: $w_0=1, w_2=1-y^2$ and $w_3=(1-y)^2(1+2y)$, to which only the dimension 6 and 8 condensates contribute at leading order, and extracts the values for $s_{\mathrm{min}}, \alpha_s$ and the OPE and DV parameters by fitting in a window $s_{\mathrm{min}} < s_0 < m_\tau^2$ ,
       \item uses V and A data, and checks the results of the fits against the spectral functions,
  \item checks the WSRs. In our case, these sum rules are satisfied.
\end{itemize}
The summary is that the results we find from the fits are consistent in all cases, whether we use only the $V$ channel or the $V$ channel together with the $A$ channel. They are also consistent whether we use just the $w_0$ weight,  the combination of $w_0$ with $w_2$, or the 3 weights $w_{0,2,3}$ together. In particular, we obtain \cite{Boito:2014sta}:
\begin{eqnarray}
\label{results}
\alpha_s(m_\tau^2)&=&0.296\pm 0.010\quad (\mathrm{FOPT}) ,\nonumber \\
\alpha_s(m_\tau^2)&=&0.310\pm 0.014\quad (\mathrm{CIPT}) ,
\end{eqnarray}
depending on whether we use fixed-order perturbation theory (FOPT) or contour-improved perturbation theory (CIPT) for the
integrated D=0 OPE series. In the traditional analysis \cite{ALEPH13}, one finds central values which are larger than those in  (\ref{results}) by $\sim +0.03$, and with errors which are about half the size of our errors.   We emphasize that the difference in the size of the errors is due to the neglect of systematics in the traditional analysis.

\begin{figure}[t]
\includegraphics*[width=3.8cm]{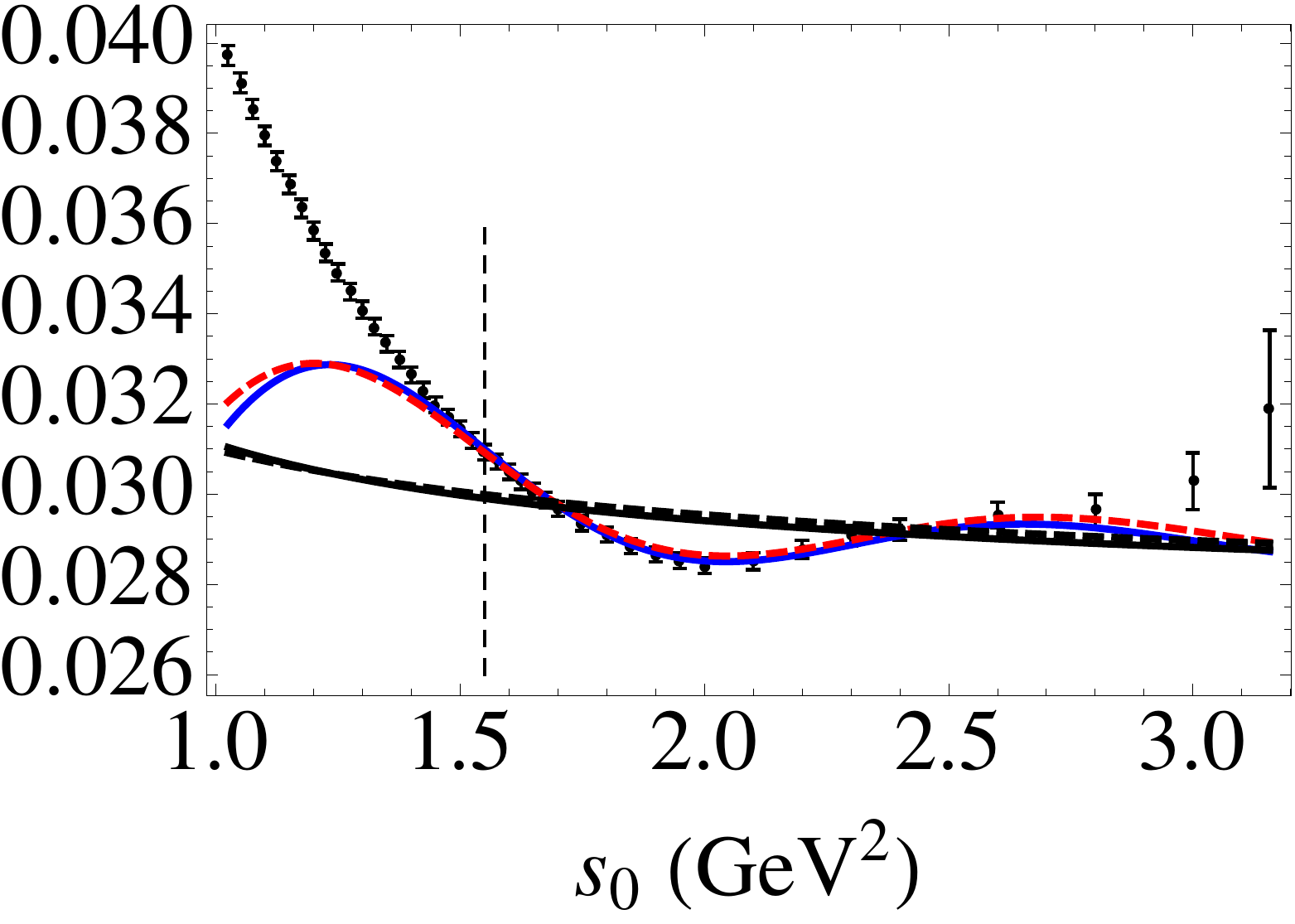}
\includegraphics*[width=3.8cm]{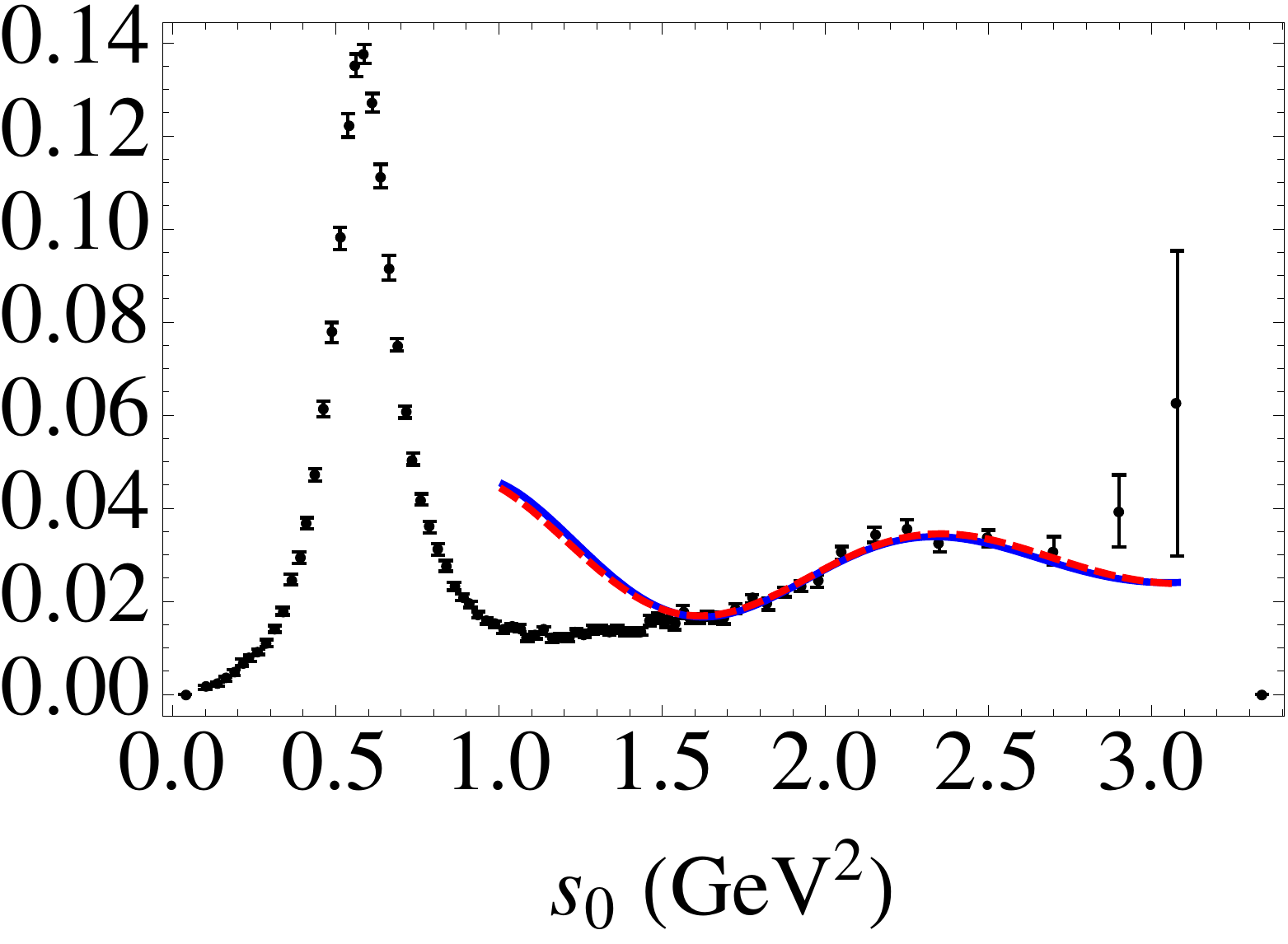}
\includegraphics*[width=3.8cm]{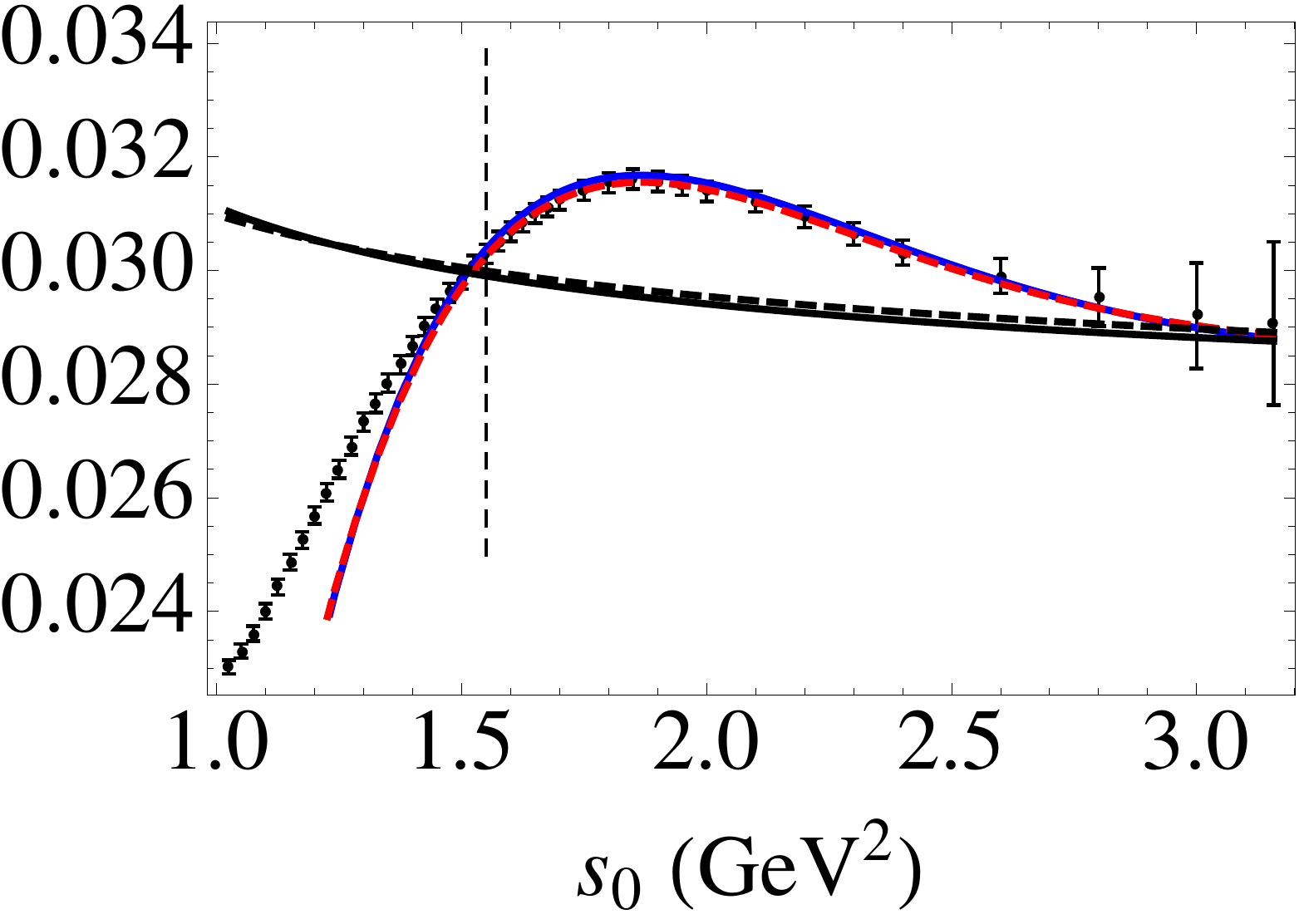}
\includegraphics*[width=3.8cm]{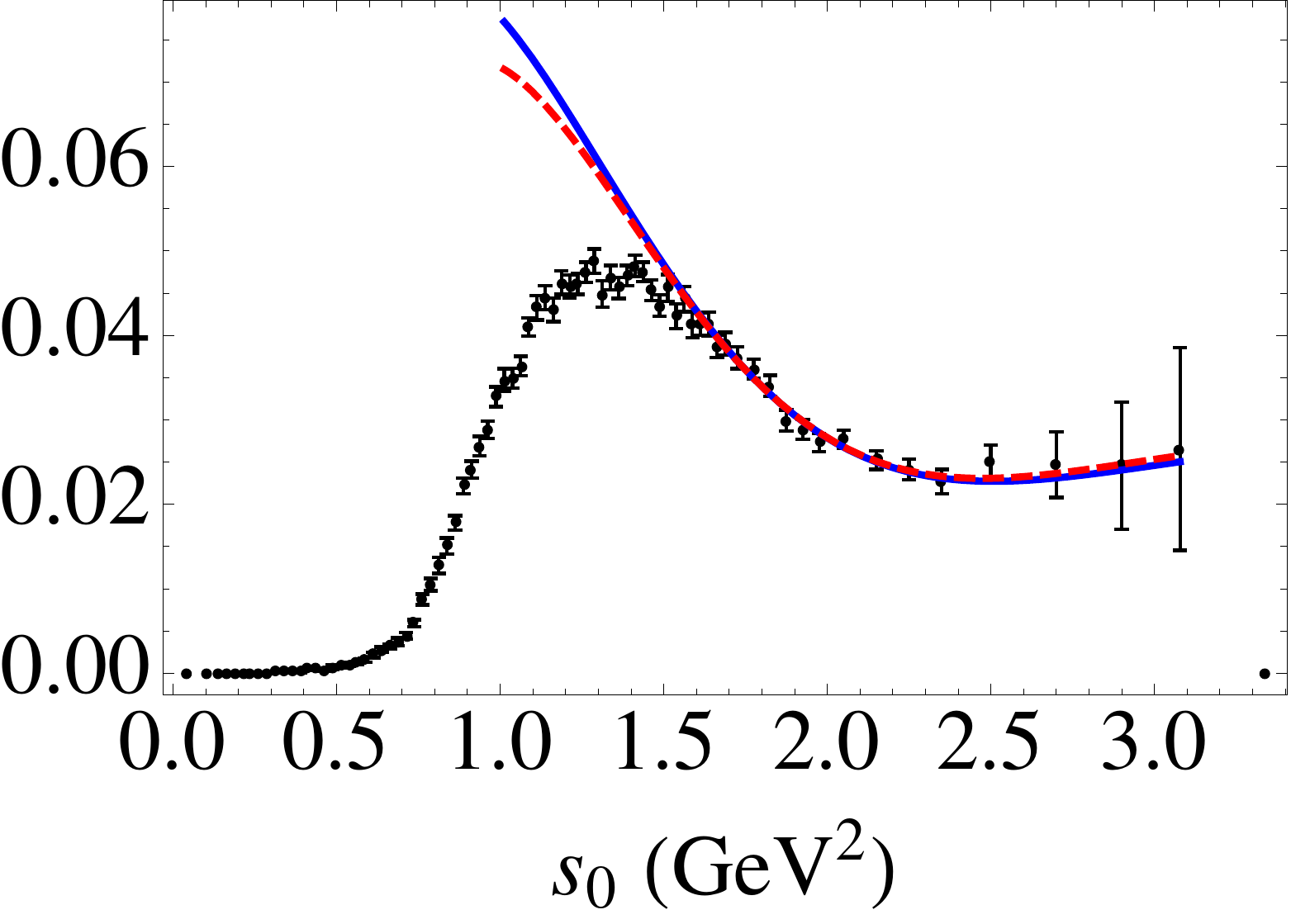}
\vspace{-.2cm}
\caption{\label{fig.2} Left panels: comparison of
$I^{(w_0)}_{\rm exp}(s_0)$ and $I^{(w_0)}_{\rm th}(s_0)$ for the combined V and A channel fit with
$s_{\rm min}=1.55\ {\rm GeV}^2$.
Right panels: comparison of the theoretical spectral function resulting
from this fit with the experimental results ($V$ top, $A$ bottom).
CIPT fits are shown in red (dashed) and FOPT in blue (solid).
The (much flatter) black curves on the left represent the OPE parts of the fits.
The vertical dashed line indicates the location of $s_{\rm min}$, determined by the fit.}
\end{figure}

Let us now see how well our fits represent the data. Figure \ref{fig.2}, for example,  shows the result of the fit using the $V$ and $A$ channels and the three weights $w_{0,2,3}$ together,  in the window $s_{\mathrm{min}}=1.55\ \mathrm{GeV}^2\leq s_0 \leq m^2_{\tau}$.
The agreement between the results of the fits and the experimental spectral functions in the region of the fit can be
seen in the right panels of Fig. \ref{fig.2}. The left panels show the comparison of the $w_0$-weighted finite energy sum rule for the $V$ and $A$ channels  with the result of the fit. The black curves show the result with DVs turned off. See Ref.~\cite{Boito:2014sta} for  further details.

\begin{figure}[t]
\begin{center}
\includegraphics*[width=5cm]{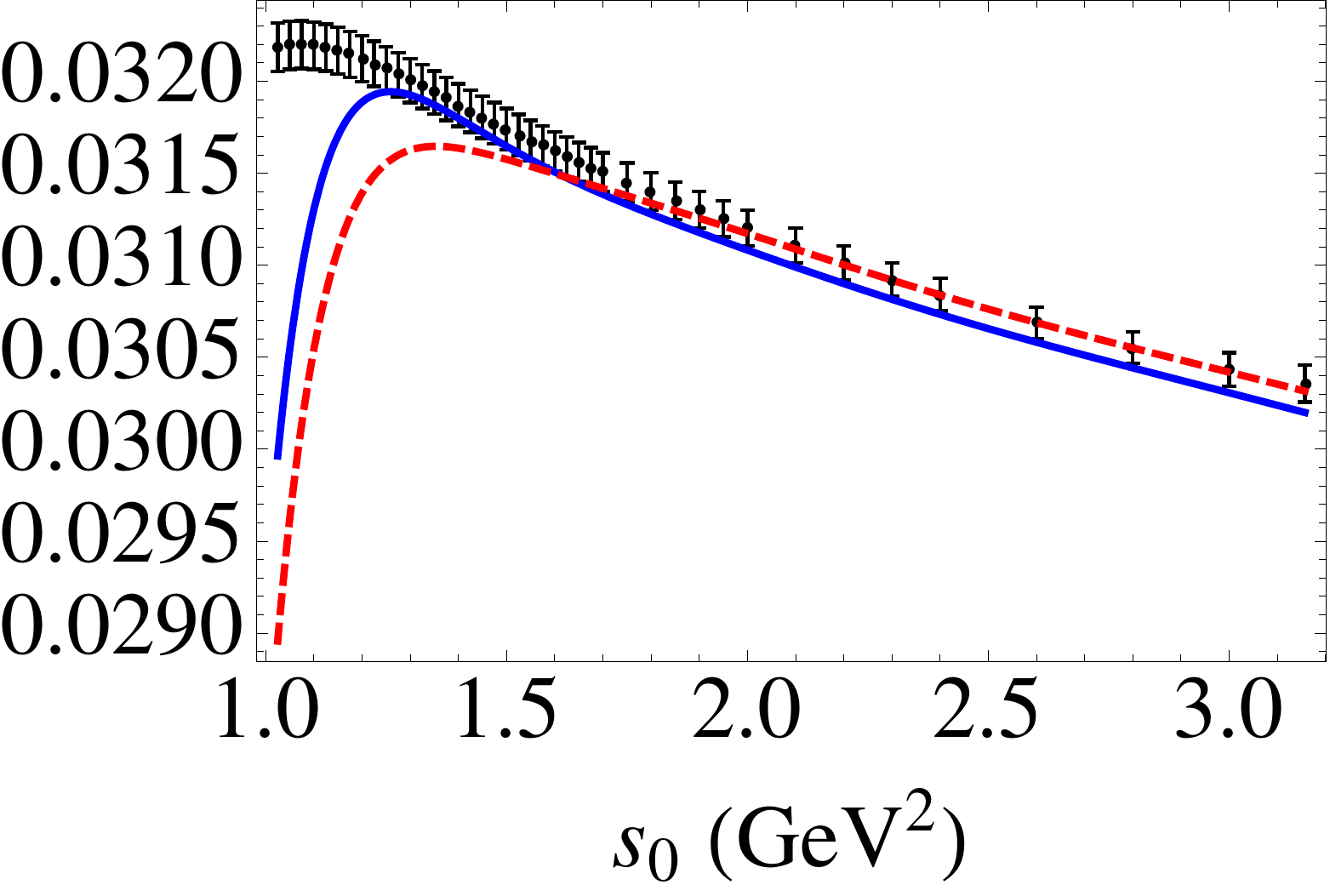}
\end{center}
\vspace{-.8cm}
\caption{\label{fig.3}  $w_3$-weighted FESR  for the $V+A$ combination, as a function of $s_0$. As before, the theory curves use $s_{\rm min}=1.55$~{\rm GeV}$^2$; CIPT (red, dashed) and FOPT (blue, solid).}
\end{figure}

Figure \ref{fig.3} shows the comparison of the results of our fits with the $w_3$-weighted FESR for the $V+A$ combination. This comparison has been considered in the past as a very important confirmation for the results obtained in traditional analysis \cite{Barate}. We now see, however, that our results also pass this test. We conclude, therefore, that this test is actually not able to discriminate between these two different sets of results.

\begin{figure}[t]
\includegraphics*[width=3.5cm]{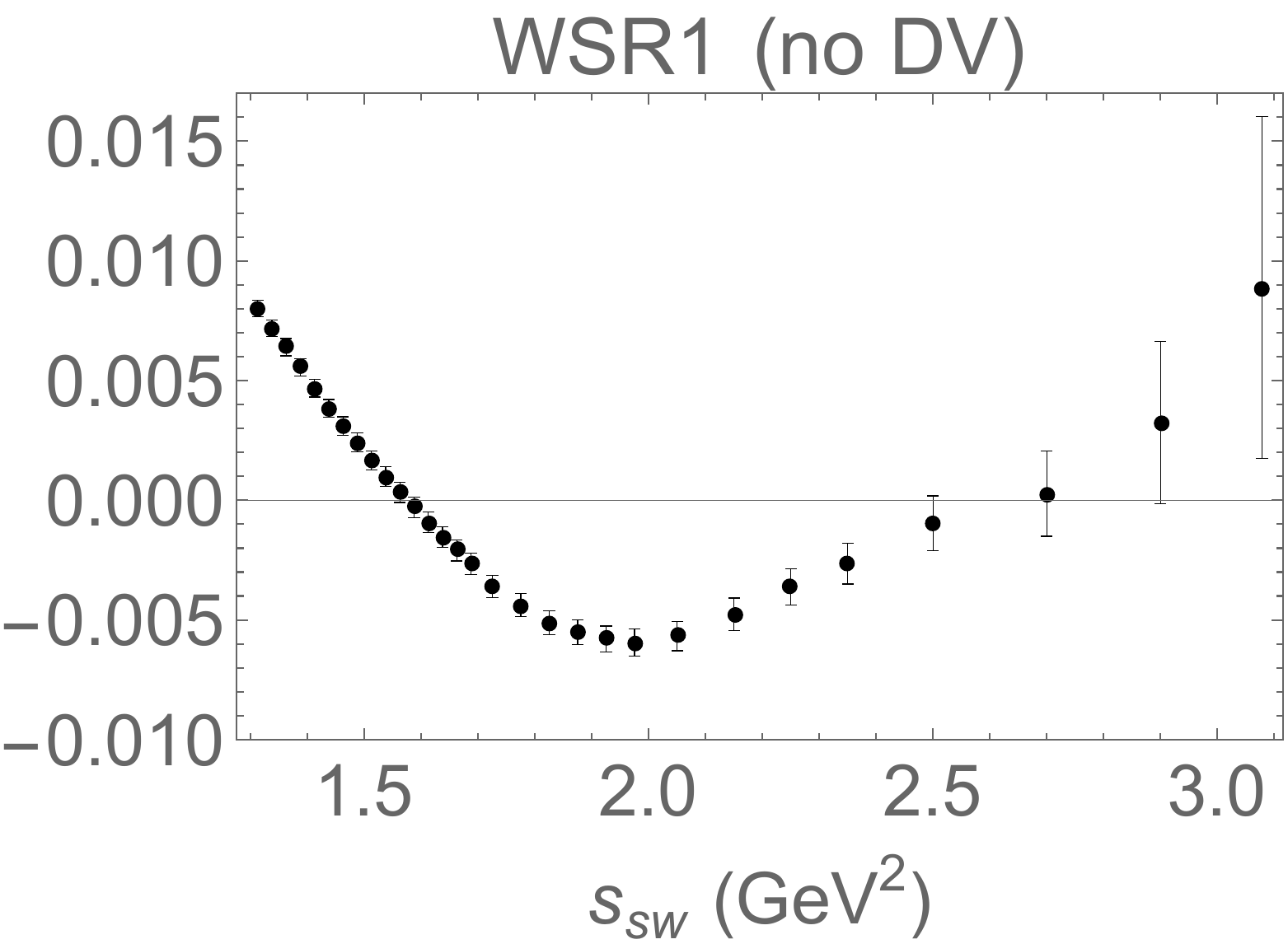}
\includegraphics*[width=3.5cm]{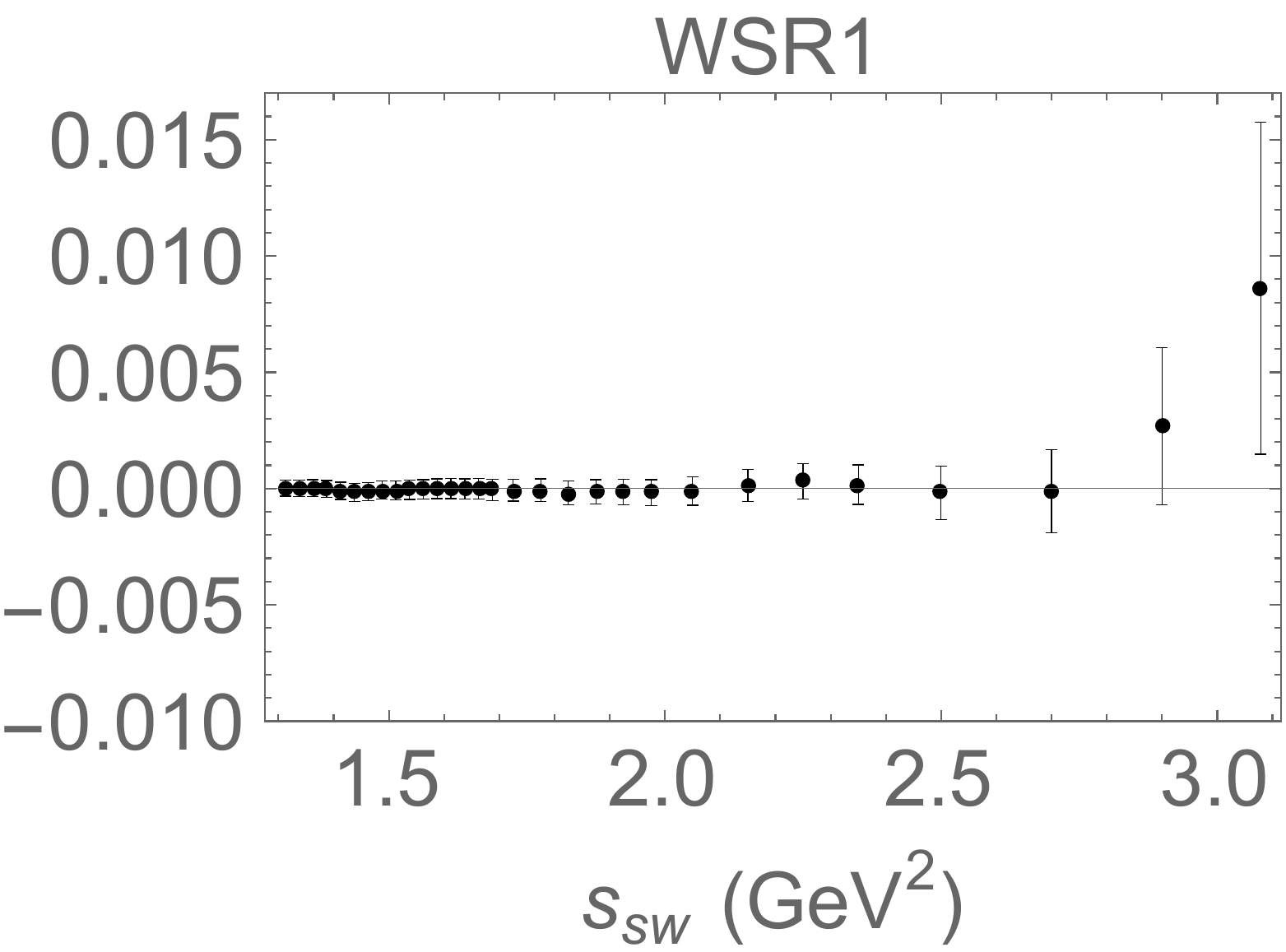}
\vspace{-.2cm}
\caption{\label{WSR} The first Weinberg sum rule, without DVs (left
panel) and with DVs (right panel), both in {\rm GeV}$^2$.
Data have been used for $s<s_{\rm sw}$, while the DV ansatz
with parameter values obtained from the $s_{\rm min}=1.55$~{\rm GeV}$^2$ fit
has been used for $s>s_{\rm sw}$. The figures shown use CIPT fits, but FOPT fit results are identical.}
\end{figure}

A test which does discriminate between the two strategies is, e.g., the verification of the WSRs. Figure \ref{WSR} shows the result for the first of those sum rules:
\begin{equation}
\label{WSRdefs}
\int_0^\infty ds\,\left(\rho^{(1)}_V(s)-\rho^{(1)}_A(s)\right)-2f_\pi^2=0\ .
\end{equation}
As the left panel shows, the sum rule is not satisfied when DVs are neglected. This is precisely the result obtained in the traditional analysis, cf. Fig. 8 in \cite{Barate}, for example, except that we have now employed the latest ALEPH data \cite{ALEPH13}. In contrast, once DVs are included, the sum rule is well satisfied (cf. the right panel). The situation with the second Weinberg sum rule is similar.

\begin{figure}[t]
\includegraphics*[width=3.8cm]{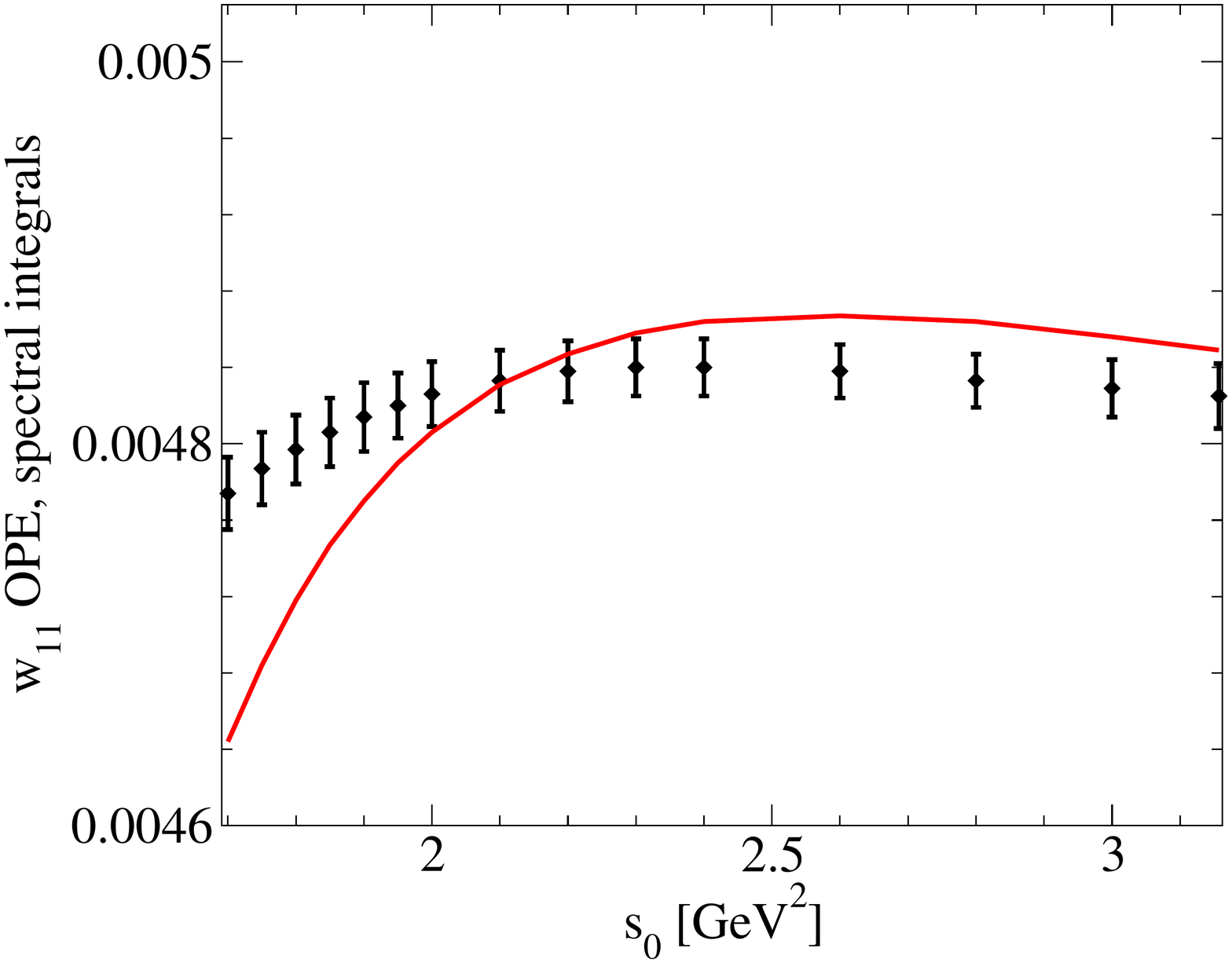}
\includegraphics*[width=3.8cm]{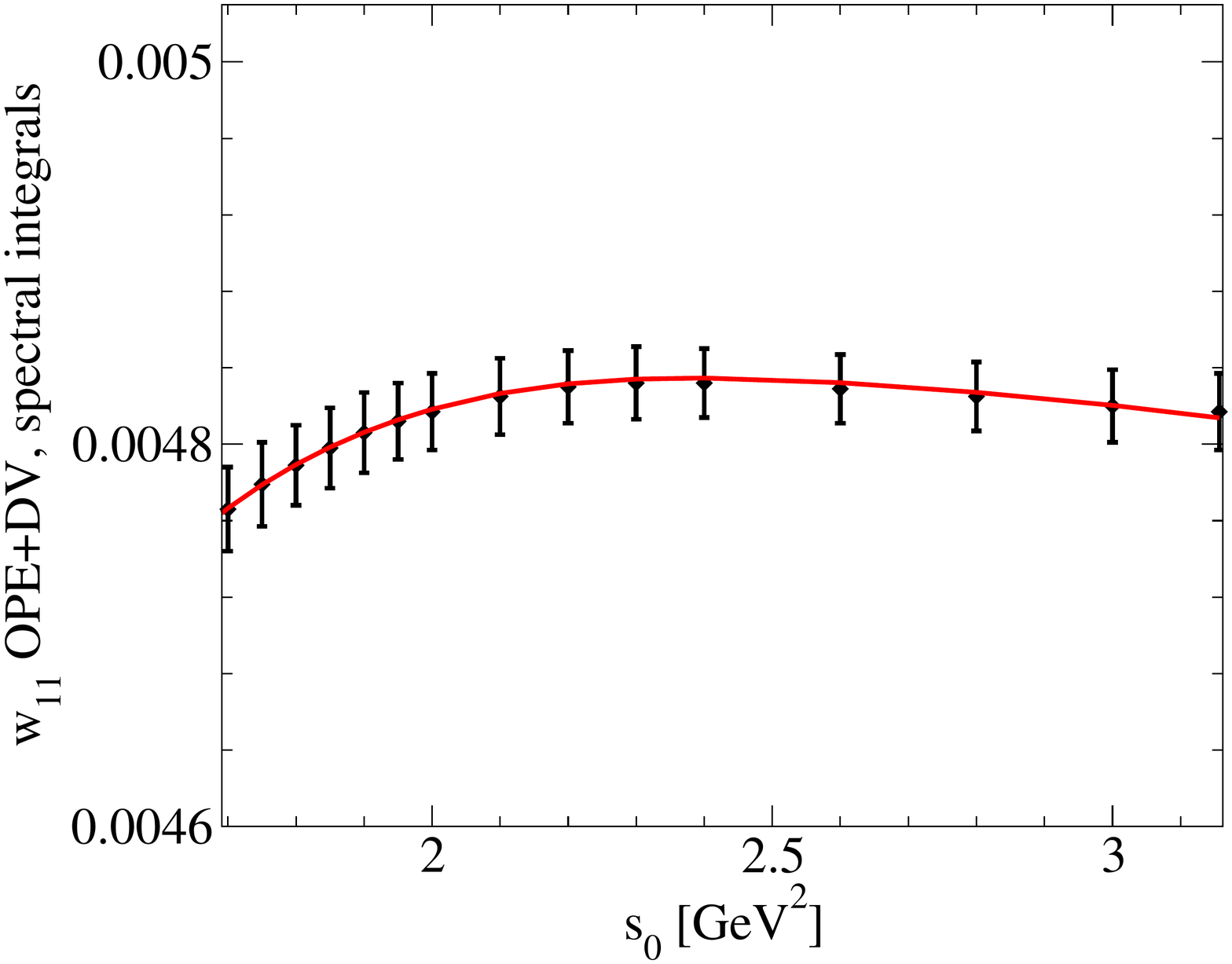}
\caption{\label{fig.4} Comparison of the $s_0$ dependence of the
$w_{11}$ $V+A$ spectral integrals and that
of the corresponding OPE integrals in the traditional analysis (left panel)  \cite{ALEPH13} and the same comparison with the results of our fit (right panel).}
\vspace*{-.2cm}
\end{figure}

One could perhaps argue that, since  the WSRs are blind to perturbation theory, the relevance of satisfying these SRs as a test for the correctness in the determination of $\alpha_s$  is rather limited. In order to fill this gap, we may compare the results of our fit with the results of the traditional analysis for the \emph{same} weights employed in the fit of the traditional analysis. One example of this comparison is shown in Fig. \ref{fig.4}.

As one can see on the left panel in this figure, even though $w_{11}$ is one of the weights employed in the traditional fit, the $s_0$ dependence of the data is rather different from that of the fit.

 In contrast, as shown in the right panel, the results corresponding to our fit reproduce the data very well, even though the moment with this weight was \emph{not} part of our fit. For other similar examples, see 
 Ref.~\cite{Boito:2014sta}.

Averaging the values of Eq. (\ref{results})  with
our previously obtained values based on  the OPAL data \cite{alphas2}, we find as our best estimate for $\alpha_s$ from $\tau$ decays,
\begin{eqnarray}
\label{results2}
\alpha_s(m_\tau^2)&=&0.303\pm 0.009\quad (\mathrm{FOPT}) ,\nonumber \\
\alpha_s(m_\tau^2)&=&0.319\pm 0.012\quad (\mathrm{CIPT}).
\end{eqnarray}

In summary, the traditional analysis \cite{DP1992}, which was still used in Ref.~\cite{ALEPH13} to analyze the revised ALEPH data,  suffers from serious self-consistency problems and, thus, should be abandoned. If the future goal is a more accurate determination of $\alpha_s$ from $\tau$ decays, this will surely require not only still better data (like those of Belle and Babar) but also a better understanding of the OPE and DVs than those available at present. A potential step in this direction has been taken in Ref.~\cite{CGP14}. We have, however, conclusively shown that the traditional strategy, with its neglect of DVs and arbitrary truncation of the OPE, should no longer be followed, as already emphasized in Refs.~\cite{alphas1,alphas2,Boito:2012mr}.

\vspace{.2cm}
DB was supported by DFG (Germany) and the A.v. Humboldt Foundation. MG and JO are supported by the DOE (USA). KM is supported by NSERC (Canada) and SP is supported by FPA2011-25948, CPAN (CSD2007-00042) and  2014~SGR~1450 (Spain).









\end{document}